\documentclass[12pt]{article}
\usepackage{amsmath,amssymb,amsfonts,epsf}
\def\ci{\mathrm{i}}
\def\ud{\mathrm{d}}

\newcommand{\expp}{\mathrm{e}}
\renewcommand{\Re}{\operatorname{Re}}

\renewcommand{\v}{\boldsymbol v}
\newcommand{\F}{\boldsymbol F}
\newcommand{\s}{\boldsymbol s}
\renewcommand{\u}{\boldsymbol u}

\begin{document}
~\vspace{3.0cm}

\centerline{\LARGE Semiclassical Universality of}
\vspace{0.3cm}
\centerline{\LARGE Parametric Spectral Correlations}
\vspace{2.0cm}
\centerline{\large 
Jack Kuipers\footnote{E-mail: jack.kuipers@bristol.ac.uk}
and Martin Sieber\footnote{E-mail: m.sieber@bristol.ac.uk}}
\vspace{0.5cm}

\centerline{School of Mathematics, University of Bristol,
Bristol BS8\,1TW, UK}

\vspace{5.0cm}
\centerline{\bf Abstract}
\vspace{0.5cm}
We consider quantum systems with a chaotic classical limit
that depend on an external parameter, and study correlations
between the spectra at different parameter values. In particular,
we consider the parametric spectral form factor $K(\tau,x)$
which depends on a scaled parameter difference $x$.
For parameter variations that do not change the symmetry of the
system we show by using semiclassical periodic orbit expansions
that the small $\tau$ expansion of the form factor agrees with
Random Matrix Theory for systems with and without time reversal
symmetry.
\vspace{2.5cm}

\noindent PACS numbers: \\
\noindent 03.65.Sq ~ Semiclassical theories and applications. \\
\noindent 05.45.Mt ~ Semiclassical chaos (``quantum chaos'').

\clearpage

\section{Introduction}

One of the characteristic features of quantum systems with underlying 
chaotic dynamics lies in statistical fluctuations of their spectra.
If the energy levels are scaled such that their mean separation is one
then the statistical distribution of the levels of individual quantum 
chaotic systems are found to be universal in the semiclassical limit 
$\hbar \rightarrow 0$ and to agree with those of eigenvalues of random
matrices \cite{BGS84}. The appropriate ensembles of random matrices
depend only on symmetries of the system. Systems with or without
time reversal symmetry (TRS) are described by the Gaussian Orthogonal
Ensemble (GOE) and the Gaussian Unitary Ensemble (GUE) (in the
absence of half-integer spin and other symmetries, which we will
assume in the following).

Universality can be observed, however, not only in the spectrum of an
individual quantum system, but also in the way in which the spectrum
changes due to an external perturbation. In the following we consider
quantum systems that depend on an external parameter whose alteration
does not change the symmetry of the system and for which the
classical dynamics is chaotic for any parameter value.
Correlations between the spectra of these systems at different
parameter values are also found to be universal functions of the
parameter difference provided that the parameter is scaled in
an appropriate way \cite{SA93a,SA93c}. The universal correlation
functions again agree with those of Random Matrix Theory (RMT) and have
been derived for the GOE and the GUE. The early developments of parametric
spectral correlations are reviewed in \cite{GGW98}.

One main approach to understanding the connection between quantum chaos
and RMT has been the application of semiclassical
methods. One convenient quantity to characterize correlations
within the spectrum of an individual quantum system is the
spectral form factor $K(\tau)$, the Fourier transform of the
two-point correlation function of the density of states.
Within the semiclassical approximation it is expressed by
a double sum over the periodic orbits of the classical system.
An evaluation of this double sum in the diagonal approximation,
which pairs orbits with themselves or their time reverse,
yields the first term in the small $\tau$ expansion of $K(\tau)$
in agreement with RMT \cite{HO84,Ber85}. Higher order terms
are due to pairs of correlated periodic orbits. The origin
of these correlations was identified and the next-order
term obtained for uniformly hyperbolic systems in \cite{SR01,Sie02}.
The contributions of all correlated periodic orbit pairs
that contribute to the small $\tau$ expansion of $K(\tau)$ 
were evaluated and summed up by combinatorial methods
in \cite{MHBHA04,MHBHA05,Mue05}. 
Similar methods have been applied since to derive off-diagonal
terms, for example, for the conductance \cite{RS02,HMBH06}, the
shot-noise \cite{BHMH06} and the GOE-GUE transition \cite{NS06}. 

For parametric spectral correlation functions agreement with 
RMT has been obtained within the diagonal approximation 
\cite{GSBSWZ91,BK96,OLM98}. 
In this article we go beyond the diagonal approximation and
derive off-diagonal terms for the parametric spectral form factor.
For systems without time reversal symmetry we derive all terms in
the small $\tau$-expansion in a closed form. In the
GOE case the method allows the calculation of  arbitrarily many
terms in the expansion. We compare the expansion up to $\tau^7$
with the result of RMT and find exact agreement.
One main reason for the universal result is that in the semiclassical
limit $\hbar \rightarrow 0$ the relevant quantum fluctuations
are due to very small parameter variations on the classical scale.
One assumption is that the parameter dependence is in some sense
typical. Specifically, we assume that the derivatives
of the actions of very long periodic orbits with respect to the
parameter have a Gaussian distribution \cite{BK96,OLM98}. This
excludes, for example, rank-one perturbations for which off-diagonal
terms were calculated in \cite{Sie00}.

In section~2 we introduce the parametric spectral form factor
and in section~3 we state results of Random Matrix Theory
for it. In section~4 we consider its semiclassical approximation
and the diagonal approximation, while in section~5 we derive the
off-diagonal terms. For systems without time reversal symmetry
the expansion is summed in section~6, and section~7 contains
our conclusions. 

While writing up our paper the preprint \cite{NBMSHH06},
which is closely related to our work, appeared on the archive. 
Nagao et al. investigate parametric correlations
that depend on a magnetic field difference, and obtain
the universal results for the GUE case and the GOE-GUE
transition by periodic orbit expansions. Our work is
complementary in that we treat arbitrary parameters and
consider also the GOE case.

\section{The parametric spectral form factor}

One way to characterize fluctuations in quantum spectra
is to consider correlation functions of the density of states.
For parametric correlations the density of states depends on
the energy $E$ as well as on a parameter $X$, and in the
semiclassical regime it can be written
as the sum of a mean part and an oscillatory part
\begin{equation}
d(E,X) = \sum_n \delta(E - E_n(X)) \approx \bar{d}(E,X) + 
d^{\text{osc}}(E,X) \; ,
\end{equation}
where $E_n(X)$, $n=1,2,\ldots$, is the $n$-th energy level as a function
of the parameter $X$. The mean density of states in an $f$-dimensional system
is given by $\bar{d}(E,X) \sim \Omega(E,X)/(2 \pi \hbar)^f$
in the semiclassical limit $\hbar \rightarrow 0$.
$\Omega(E,X)$ is the volume of the surface of constant energy in
phase space at energy $E$ and parameter $X$.

In order to obtain a universal parametric spectral correlation function
one has to perform two unfoldings, one in energy $E$ and one in the
parameter $X$ of the system. A new energy parameter is defined by
\begin{equation} \label{unfolde}
\tilde{E} = \bar{N}(E,X) \; ,
\end{equation}
where $\bar{N}(E,X)$ is the mean part of the spectral staircase 
$N(E,X) = \int_{-\infty}^E \ud E' \; d(E',X)$. 
In terms of the new energy $\tilde{E}$ the density of states has a mean
value of one. The spectral statistics are evaluated in the semiclassical
limit in an interval $\Delta \tilde{E}$ that is classically small but
contains a large number of energy levels, i.e. it satisfies $\tilde{E} \gg
\Delta \tilde{E} \gg 1$.

A new parameter $\tilde{X}$ is introduced by
\cite{SA93a,LS99}
\begin{equation} \label{unfoldx}
\tilde{X} = \int_{X_0}^X d X' \; \sigma(X') \; , \qquad \sigma(X') = 
\sqrt{\langle v_n(X')^2 \rangle} \; ,
\end{equation}
where $v_n(X) = \partial \tilde{E}_n / \partial X$ are the level velocities
and the average is performed over the 
levels in the interval $\Delta \tilde{E}$.
$X_0$ is an arbitrary parameter value at which $\tilde{X}=0$.
In terms of the new parameter $\tilde{X}$ the level velocities have a
unit variance.

We may then define the universal two-point correlation function by
\begin{equation}
R_2(\eta,x) = \left\langle \tilde{d}^{\text{osc}}
\left(\tilde{E}+\frac{\eta}{2},\tilde{X}+\frac{x}{2}\right) \;
\tilde{d}^{\text{osc}}
\left(\tilde{E}-\frac{\eta}{2},\tilde{X}-\frac{x}{2}\right)
\right\rangle_{\tilde{E},\tilde{X}} \; ,
\end{equation}
where $\tilde{d}(\tilde{E},\tilde{X})$ is the density of states
of the unfolded spectrum, and 
the average is performed over the energy interval $\Delta \tilde{E}$
as well as over a parameter interval $\Delta \tilde{X}$. 
The relation to the original density of states is given by
\begin{equation}
\tilde{d}(\tilde{E},\tilde{X}) = \frac{\partial N(E,X)}{\partial E} 
\frac{\partial E}{\partial \tilde{E}} = \frac{d(E,X)}{\bar{d}(E,X)} \; ,
\end{equation}
and in the semiclassical limit we find that
\begin{equation} \label{r2}
R_2(\eta,x) \sim \frac{\left\langle
d^\text{osc}\left(E + \frac{\eta}{2 \bar{d}} + \frac{x \rho}{2 \sigma}
,X + \frac{x}{2 \sigma} \right) \;
d^\text{osc}\left(E - \frac{\eta}{2 \bar{d}} - \frac{x \rho}{2 \sigma},
X - \frac{x}{2 \sigma} \right) \;
\right\rangle_{E,X}}{\bar{d}(E,X)^2} \; .
\end{equation}
Equation~(\ref{r2}) has been obtained by linearizing the unfolding 
equations~(\ref{unfolde}) and~(\ref{unfoldx}), because $x$ and $\eta$ correspond to
small changes on the classical scale, since $\bar{d}$ is of the order
$\hbar^{-f}$ and $\sigma=\sigma(X)$ is of the order of $\hbar^{-(f+1)/2}$
(see equation (\ref{semisigma}) later).
The term $x/2 \rho \sigma$ takes account of the change of the energy when $X$
is changed while keeping $\tilde{E}$ fixed
\begin{equation}
\rho = \left. \frac{\partial E}{\partial X} \right|_{\tilde{E}} =
- \frac{\partial \bar{N} / \partial X}{\partial \bar{N} / \partial E} \; .
\end{equation} 

In the following, we will consider the parametric spectral form factor
which is obtained by a Fourier transform of the parametric two-point
correlation function
\begin{equation} \label{defform}
K(\tau,x) = \int_{-\infty}^\infty R_2(\eta,x) \, e^{-2 \pi i \eta \tau}
\; d \eta \; .
\end{equation}

\section{Results from Random Matrix Theory}

The parametric two-point correlation function $R_2(\eta,x)$ has been derived in
the context of disordered systems 
\cite{SA93a,SA93b} for the GUE and the GOE. For the GUE case it is given by
\begin{equation}
R_2^{\text{GUE}}(\eta,x) = \frac{1}{2} \int_{-1}^1 \ud \lambda \int_1^\infty \ud \lambda_1 \; 
\cos(\pi\omega(\lambda_1-\lambda)) \expp^{-\pi^2 x^2({\lambda_1}^2-\lambda^2)/2} \; .
\end{equation}
After performing the Fourier transform in~(\ref{defform}) to obtain the parametric form 
factor we arrive at
\begin{equation}
K^{\text{GUE}}(\tau,x)=\frac{1}{2} \int_{-1}^{1} \ud \lambda \int_{1}^{\infty} \ud \lambda_1 \;
\expp^{-\pi^{2}x^2({\lambda_{1}}^{2}-\lambda^{2})/2}
\left[\delta(\lambda_{1}-\lambda-2\tau)
+ \delta(\lambda_{1}-\lambda+2\tau)\right] \; .
\end{equation}
Because $\tau$ is positive and $\lambda_{1}\geq \lambda$ the second delta function
does not contribute.  From the first delta function we get the relation
$2\tau=\lambda_{1}-\lambda$. In the case $\tau<1$, which we consider in the 
following, the domain of integration for $\lambda_1$ is reduced to
$1\leq \lambda_{1}\leq1+2\tau$, and we obtain
\begin{equation} \label{kgue1}
K^{\text{GUE}}(\tau,x)= \frac{1}{2} \int_{1}^{1+2\tau} \ud \lambda_1 \; 
\expp^{2\pi^{2}x^{2}\tau(\tau-\lambda_1)}
=\frac{\sinh(2\pi^{2}x^{2}\tau^{2})}{2\pi^{2}x^{2}\tau}\expp^{- 2 \pi^2 x^2 \tau}
\; , \quad \tau < 1 \; .
\end{equation}
For comparison with the semiclassical expansion we expand the sinh-function
and define $B=2 \pi^2 x^2/\kappa$ where $\kappa=1$ and $2$ for the GUE and GOE
cases, respectively.
\begin{equation} \label{kgue2}
K^{\text{GUE}}(\tau,x) = \expp^{-B \tau} \sum_{n=0}^\infty 
\frac{B^{2 n} \tau^{4 n + 1}}{(2 n + 1)!} \; .
\end{equation}

The parametric correlation function for the GOE case is given by a triple integral
\begin{align}
R_2^{\text{GOE}}(\eta,x)&= \int_{-1}^{1} \ud \lambda \int_{1}^{\infty} \ud \lambda_1 \int_{1}^{\infty} 
\ud \lambda_2 \; \cos(\pi\omega(\lambda-\lambda_{1}\lambda_{2}))
\frac{(1-\lambda^{2})(\lambda-\lambda_{1}\lambda_{2})^{2}}{
(2\lambda\lambda_{1}\lambda_{2}-\lambda^{2}-{\lambda_{1}}^{2}-{\lambda_{2}}^{2}+1)^{2}}
\notag \\
& \qquad \qquad \times
\expp^{-\pi^{2}x^2 (2{\lambda_{1}}^{2}{\lambda_{2}}^{2}
-\lambda^{2}-{\lambda_{1}}^{2}-{\lambda_{2}}^{2}+1)/4} \; .
\end{align}
Evaluating the Fourier transform to obtain the parametric form factor results in
a sum of two delta-functions
\begin{align}
K^{\text{GOE}}(\tau,x)&= \int_{-1}^{1} \ud \lambda \int_{1}^{\infty} \ud \lambda_1 
\int_{1}^{\infty} \ud \lambda_2 
\frac{(1-\lambda^{2})(\lambda-\lambda_{1}\lambda_{2})^{2}}{
(2\lambda\lambda_{1}\lambda_{2}-\lambda^{2}-{\lambda_{1}}^{2}-{\lambda_{2}}^{2}+1)^{2}}
\notag \\ 
& \qquad \times
\expp^{-\pi^{2}x^2(2{\lambda_{1}}^{2}{\lambda_{2}}^{2}-\lambda^{2}-
{\lambda_{1}}^{2}-{\lambda_{2}}^{2}+1)/4}
\left[\delta(\lambda-\lambda_{1}\lambda_{2}-2\tau)
     +\delta(\lambda-\lambda_{1}\lambda_{2}+2\tau)\right] \; .
\end{align}
Because $\tau$ is positive and $\lambda_{1}\lambda_{2}\geq \lambda$ only the second
delta function contributes, giving the relation $\lambda=\lambda_{1}\lambda_{2}-2\tau$. 
As we are again considering the case when $\tau<1$ our domain of integration for the
other two variables is given by $1\leq \lambda_{1}\leq1+2\tau$ and
$1\leq \lambda_{2}\leq\frac{1+2\tau}{\lambda_{1}}$.  When we perform the integral
over $\lambda$ we are left with
\begin{align}
K^{\text{GOE}}(\tau,x)&=
\int_{1}^{1+2\tau} \ud \lambda_1 \int_{1}^{\frac{1+2\tau}{\lambda_{1}}}  \ud \lambda_2 \;
\frac{4\tau^{2}(1-{\lambda_{1}}^{2}{\lambda_{2}}^{2}+4\tau\lambda_{1}\lambda_{2}-4\tau^{2})
}{(1+{\lambda_{1}}^{2}{\lambda_{2}}^{2}-{\lambda_{1}}^{2}-{\lambda_{2}}^{2}-4\tau^{2})^{2}}
\notag \\ & \qquad \qquad \times 
\expp^{-\pi^{2}x^{2}(1+{\lambda_{1}}^{2}{\lambda_{2}}^{2}-{\lambda_{1}}^{2}-{\lambda_{2}}^{2}
+4\tau\lambda_{1}\lambda_{2}-4\tau^{2})/4} \; .
\end{align}
In order to evaluate this integral as a series in $\tau$ it is useful to remove the $\tau$ 
dependence from the limits. This is done by changing the integration variables using
$\lambda_1=1+ \tau y_1$ and $\lambda_1 \lambda_2 = 1 + \tau y_2$. Then the expansion of 
the parametric form factor is obtained by expanding the integrand for small values
of $\tau$. 
\begin{align}
K^{\text{GOE}}(\tau,x)& = \int_{0}^{2} \ud y_1 \int_{y_{1}}^{2} \ud y_2 \; \left\{
\frac{2-y_{2}}{2(1-y_{1}y_{2}+{y_{1}}^{2})^{2}}\tau + 
\left[\frac{y_{1}(y_{2}-2)(4-y_{1}y_{2}+2{y_{1}}^{2}-{y_{2}}^{2})}{
2(1-y_{1}y_{2}+{y_{1}}^{2})^{3}}\right. \right. \notag \\
& \qquad \qquad \left. \left. + \frac{(2-y_{2})(2+2\pi^{2}x^{2}-6y_{1}-y_{2})}{
4(1-y_{1}y_{2}+{y_{1}}^{2})^{2}}\right]\tau^{2} + \ldots \right\} \; .
\end{align}
Using Maple we performed this expansion up to seventh order and evaluated
the integrals with the following result
\begin{align}
K^{\text{GOE}}(\tau,x) & = 2\tau-(2\pi^{2}x^{2}+2)\tau^{2}+(\pi^{4}x^{4}+2)\tau^{3}-
\left(\frac{\pi^{6}x^{6}}{3}-\pi^{4}x^{4}+\frac{8}{3}\right)\tau^{4} \notag \\
&+\left(\frac{\pi^{8}x^{8}}{12}-\frac{2\pi^{6}x^{6}}{3}+\frac{2\pi^{4}x^{4}}{3}+4\right)\tau^{5}
\notag \\
&-\left(\frac{\pi^{10}x^{10}}{60}-\frac{\pi^{8}x^{8}}{4}+\pi^{6}x^{6}+
\frac{\pi^{4}x^{4}}{3}+\frac{32}{5}\right)\tau^{6}
\notag \\ & +\left(\frac{\pi^{12}x^{12}}{360}-\frac{\pi^{10}x^{10}}{15}+
\frac{7\pi^{8}x^{8}}{12}-\frac{2\pi^{6}x^{6}}{15}+\frac{\pi^{4}x^{4}}{5}
+\frac{32}{3}\right)\tau^{7} \; .
\end{align}
For comparison with the semiclassical result it is convenient to extract
an exponential factor from this expansion. 
\begin{align} \label{kgoe}
K^{\text{GOE}}(\tau,x) & = \expp^{-B\tau} \left[ 2 \tau - 2\tau^2 - (2B-2) \tau^3 +
\left(2 B - \frac{8}{3} \right) \tau^4 \right.
\notag \\ &
+ \left( \frac{5B^2}{3} - \frac{8B}{3} + 4 \right) \tau^5-
\left( \frac{5B^2}{3} - 4 B + \frac{32}{5} \right) \tau^6 
\notag \\ &
\left. - \left( \frac{41 B^3}{45} - \frac{11 B^2}{5} + \frac{32 B}{5}-
\frac{32}{3} \right) \tau^7 + \ldots \right] \; ,
\end{align}
where $B$ has been defined after equation~(\ref{kgue1}).

\section{Semiclassical approximation}\label{diagpara}

In this section we derive a semiclassical expression for the
parametric spectral form factor. It is closely related to 
semiclassical approximations for the parametric two-point
correlation function of the density of states \cite{BK96,OLM98}.
We start by expressing the density of states with the Gutzwiller
trace formula \cite{Gut71}
\begin{equation} \label{gutz}
d^\text{osc}(E,X) \approx \frac{1}{\pi \hbar} \Re
\sum_\gamma A_\gamma \exp\left( \frac{\ci}{\hbar} S_\gamma \right)
\; , \quad
\text{where} \quad A_\gamma = \frac{T_\gamma}{R_\gamma
\sqrt{|\det(M_\gamma - 1)|}} \expp^{-\ci \pi \mu_\gamma/2} \; .
\end{equation}
The sum runs over all periodic orbits of the system with period
$T_\gamma$, repetition number $R_\gamma$, stability matrix
$M_\gamma$ and Maslov index $\mu_\gamma$.

The action is expanded in first order in the energy difference and the 
parameter difference
\begin{equation} \label{act}
S_\gamma \left(E \pm \frac{\eta}{2 \bar{d}} \pm \frac{x \rho}{2 \sigma}
,X \pm \frac{x}{2 \sigma} \right)
\approx S_\gamma(E,X) \pm T_\gamma(E,X) \frac{\eta}{2 \bar{d}}
\pm Q_\gamma(E,X) \frac{x}{2 \sigma} \; ,
\end{equation}
where 
\begin{equation}
Q_\gamma=\left. \frac{\partial S_\gamma}{\partial X} \right|_{\tilde{E}}    
=  \rho \frac{\partial S_\gamma}{\partial E} 
 + \frac{\partial S_\gamma}{\partial X}
\end{equation}
is the parametric velocity.

After inserting (\ref{gutz}) and (\ref{act}) into (\ref{r2}) and (\ref{defform})
and evaluating the integral to leading semiclassical order, we arrive at
\begin{equation} \label{ksemi}
K(\tau,x) = \frac{1}{T_H} \left\langle \sum_{\gamma,\gamma'} 
A_\gamma A_{\gamma'}^* \expp^{\frac{ \ci(S_\gamma-S_{\gamma'})}{\hbar}}
\expp^{\frac{\ci x (Q_\gamma + Q_{\gamma'})}{2 \sigma \hbar}}
\delta \left( T - \frac{T_\gamma + T_{\gamma'}}{2} \right) \right\rangle \; ,
\end{equation}
where $T_H = 2 \pi \hbar \bar{d}(E)$ is the Heisenberg time, and $\tau=T/T_H$.
Terms which have a sum of the actions in the exponent have been neglected,
because they average away. The semiclassical expression (\ref{ksemi})
is the quantity that we will evaluate in the following.

The diagonal approximation involves pairs of orbits that are either
identical or related by time reversal. It has the form 
\begin{equation}
K(\tau,x) = \frac{\kappa}{T_H} \left\langle \sum_\gamma 
|A_\gamma|^2 \expp^{\frac{\ci x Q_\gamma}{\sigma \hbar}}
\delta(T - T_\gamma) \right\rangle \; ,
\end{equation}
where $\kappa$ is 2 if the system has time reversal symmetry
and 1 if it does not. 

One main ingredient in the following semiclassical calculation
is the distribution of the parametric velocities $Q_\gamma$ in
the limit of very long periodic orbits. It has been shown that
the $Q_\gamma$ have a mean value of zero and a variance proportional
to their period \cite{GSBSWZ91}
\begin{equation} \label{qmean}
\langle Q_\gamma \rangle = 0 \; , \quad \langle Q_\gamma^2 \rangle \sim a T \; ,
\quad T \rightarrow \infty \; ,
\end{equation}
where the averages are performed over trajectories with period around $T$. 
It has been motivated that the $Q_\gamma$ have a Gaussian 
distribution for long periodic orbits (see \cite{OLM98}) and this
is the main assumption that we will use in the following.
The proportionality factor $a$ in (\ref{qmean}) is related
to the variance of the level velocities. \cite{EFKAMM95,LS99}
\begin{equation} \label{semisigma}
\sigma^2 \sim \frac{a \kappa \bar{d}}{2 \pi \hbar} \; .
\end{equation}
Using the Gaussian assumption for the distribution of the parametric
velocities and (\ref{semisigma}) we perform the average over the $Q_\gamma$,
assuming that it can be done independently, and obtain
\begin{equation} \label{qaverage}
\langle \expp^{\frac{\ci x Q_{\gamma}}{\sigma \hbar}} \rangle
= \expp^{-\frac{x^2 a T}{2 \sigma^2 \hbar^2}} = \expp^{-B T/T_H} \; ,
\end{equation}
where $B = 2 \pi^2 x^2/\kappa$ is the same quantity as defined after equation
(\ref{kgue1}). The remaining sum over periodic orbits can be evaluated 
with the Hannay-Ozorio de Almeida sum rule \cite{HO84}
\begin{equation}
\sum_\gamma |A_\gamma|^2 \delta(T - T_\gamma) \approx T \; .
\end{equation}
We find that the diagonal approximation is given by
\begin{equation} 
K(\tau) = \kappa \tau \expp^{-B \tau}
\end{equation}
in agreement with the first term in the expansion of the
random matrix results, (\ref{kgue2}) and (\ref{kgoe}).

\section{Off-diagonal contributions}

The off-diagonal terms of the parametric form factor are due to
pairs of trajectories which are correlated \cite{SR01,MHBHA05}.
In the following we briefly review the main steps in the derivation
of the semiclassical expansion of the spectral form factor
(in our notation $K(\tau,x=0)$) according to \cite{MHBHA05,Mue05}.
The correlations that are important for the expansion of the form factor for
small $\tau$ are due to close self-encounters of a periodic orbit 
in which two or more stretches of an orbit are almost identical,
possibly up to time reversal. In general, a long periodic orbit
has many of these encounter regions, and they are connected by long
parts of the orbit, the so-called ``loops''. The correlated pairs of 
orbits are almost identical along the loops, but they differ in
the way in which the loops are connected in the encounter regions. 
Correlated orbit pairs have certain ``structures'' that are 
characterized by the number of encounter regions $V$ in which
the loops are connected in a different way, the number of
involved orbit stretches $l_\alpha$ in each encounter region 
$\alpha$, and the way in which the loops are connected by these
stretches. A more accurate definition of structures can be given
by putting them in a one-to-one relation with permutation matrices
that describe the reconnections of the loops. One defines further
a vector $\v$ whose $l$-th component, $v_l$, specifies the number of
encounter regions with $l$ stretches, and the total number of
orbit stretches is denoted by $L$. Hence
\begin{equation}
V = \sum_{l \ge 2} v_l \; , \qquad L = \sum_\alpha l_\alpha = \sum_{l \ge 2} l v_l \; .
\end{equation} 
The semiclassical contribution to the form factor is evaluated
in two steps. First the summation over orbit pairs with
the same structure is evaluated by using that long periodic
orbits are uniformly distributed over the surface of constant
energy in phase space. Then the summation over the different
structures is performed which is a combinatorial problem.

In the following we present some details of this calculation.
In each encounter region $\alpha$ with $l_\alpha$ orbit stretches
one chooses a perpendicular Poincar\'e surface that is centered
on one of the stretches. The relative positions of the piercings
of the other stretches through the Poincar\'e surface are described
by coordinates along the stable and unstable manifolds. The partner
periodic orbit connects the loops that start and end at the encounter
region in a different way and the resulting contribution to the action
difference is given in the linearized approximation by
\begin{equation}
(\Delta S)_\alpha = \sum_{j=1}^{l_\alpha-1} s_{\alpha j} u_{\alpha j} \; ,
\end{equation}
where $s_{\alpha j}$, $u_{\alpha j}$, $j=1,\ldots, l_\alpha-1$ are
appropriate differences of the coordinates along the stable and
unstable manifolds. For ease of notation we discuss here the
two-dimensional case in which the coordinates $s_{\alpha j}$ and 
$u_{\alpha j}$ are scalars. 
If the Poincar\'e surface is moved along the stretches in the 
encounter region these coordinates decrease or increase,
exponentially, however their product remains constant. The length
of the encounter region is determined by requiring that all coordinates 
remain smaller than an arbitrary small constant $c$ whose exact
size is not relevant for the following calculations.

The uniform distribution of the long periodic orbits on the energy
shell is then invoked to sum over all orbit pairs with the same
structure. It is convenient to also sum over all structures with the
same vector $\v$ and express it in the form
\begin{equation} \label{sumv}
K_{\v}(\tau) = \frac{1}{T_H} \sum_{(\gamma,\gamma')}^{\text{fixed} \, \v} 
|A_\gamma|^2 e^{\ci \Delta S_\gamma/\hbar} 
\delta(T - T_\gamma) = N(\v) \kappa \tau \int d^{L-V} s \; d^{L-V} u \;
\frac{w_T(\s,\u)}{L} \expp^{\ci \s \u / \hbar}
\end{equation}
so that $K(\tau) = \kappa \tau + \sum_{\v} K_{\v}(\tau)$. $N(\v)$ is 
the number of structures with the same $\v$, and $\s$ and $\u$ are
vectors whose components are the $s_{\alpha j}$ and $u_{\alpha j}$ 
for all $\alpha$ and $j$. In (\ref{sumv}) the amplitudes and periods
of the two correlated orbits are set equal. 
$w_T(\s,\u)$ is the density of the self-encounters for a given structure
and separation coordinates $s_{\alpha j}$ and $u_{\alpha j}$. For long orbits
it is given asymptotically by
\begin{equation} \label{defw}
\frac{w_T(\s,\u)}{L} = \frac{T (T - \sum_\alpha l_\alpha t_{\text{enc}}^\alpha)^{L-1}}{
L! \Omega^{L-V} \prod_\alpha t_{\text{enc}}^\alpha } \; .
\end{equation}
The factor $1/L$ in (\ref{sumv}) takes care of an overcounting related
to the choice of an initial point of the trajectory \cite{MHBHA05}.

The integral in (\ref{sumv}) is evaluated by using
\begin{equation} \label{intenc}
\int \prod_j \ud s_{\alpha j} \; \ud u_{\alpha j} \; (t_{\text{enc}}^\alpha)^k 
\expp^{\ci \sum_j s_{\alpha j} u_{\alpha j} / \hbar} \approx
\begin{cases} 0 & \text{if} \; k=-1 \; \text{or}  \; k \ge 1 \\
              (2 \pi \hbar)^{l_\alpha-1} & \text{if} \; k=0 \end{cases} \; .
\end{equation}
Hence after expanding the numerator of (\ref{defw}) the only term that
survives is the one that contains a product of all encounter times
$t_{\text{enc}}^\alpha$ which is cancelled by the denominator.
Therefore one can replace the density $w_T$ in the integral by
\begin{equation} \label{wt}
\frac{w_T(\s,\u)}{L} \quad \Longrightarrow \quad
\frac{T}{L! \Omega^{L-V}}
\frac{(L-1)! T^{L-V-1} (-1)^V \prod_l l^{v_l}}{(L-V-1)!} \; ,
\end{equation}
where we used that $\prod_\alpha l_\alpha = \prod_l l^{v_l}$.
Inserting (\ref{wt}) into (\ref{sumv}) and evaluating the integral
with the use of (\ref{intenc}) yields the contribution from orbit pairs
with fixed $\v$,
\begin{equation}
K_{\v}(\tau) = 
\kappa \tau^{L-V+1} N(\v) \frac{(-1)^V \prod_l l^{v_l}}{L (L-V-1)!} \; .
\end{equation}
As can be seen, vectors $\v$ with the same value of $L-V+1$ contribute
to the same power of $\tau$. Hence the power series expansion of the form
factor can be obtained by summing over all $\v$ with the same $L-V+1$.
\begin{equation}  \label{semikn}
K(\tau) = \kappa \tau + \sum_{n=2}^\infty K_n \tau^n \; ,
\end{equation}
where
\begin{equation}
K_n = \frac{\kappa}{(n-2)!} \sum_{\v}^{L-V+1=n} \tilde{N}(\v) \; , \quad
\tilde{N}(\v)=(N(\v)(-1)^V \prod_l l^{v_l})/L \; .
\end{equation}
By deriving a recurrence relation for the numbers $\tilde{N}(\v)$ it is
shown that the semiclassical expansion in (\ref{semikn}) agrees with
the small $\tau$ expansion of the form factor of RMT in cases with or
without time reversal symmetry \cite{MHBHA04,MHBHA05,Mue05}.

Let us now come back to the parametric form factor $K(\tau,x)$. We
have to evaluate
\begin{equation}
K(\tau,x) = \frac{1}{T_H} \left\langle \sum_{\gamma,\gamma'} 
|A_\gamma|^2 \expp^{\frac{ \ci(S_\gamma-S_{\gamma'})}{\hbar}}
\expp^{\frac{\ci x Q_\gamma}{\sigma \hbar}}
\delta \left( T - T_\gamma\right) \right\rangle \; .
\end{equation}
It now contains an additional term involving the parametric velocities $Q_\gamma$.
As for the diagonal approximation we assume that the average over the $Q_\gamma$
can be performed independently from the actions and amplitudes of the orbits.
However, we have to look at the average (\ref{qaverage}) more carefully. It
is valid as long as there are no systematic correlations between different
parts of the {\em same} periodic orbit. Long periodic orbits have many 
encounter regions in which different stretches of an orbit are almost
identical and cannot be considered as independent or uncorrelated. 
When we calculate the contribution to the form factor of orbit pairs with
a certain structure, then we average over periodic orbits that all have
the same number and types of encounter regions. The corresponding 
correlations between different parts of an orbit have to be taken into account.
We should consider the average over the parametric velocities for the loops
and encounter regions separately. The contribution from a loop is
\begin{equation} 
\langle\expp^{\frac{\ci xQ^{loop}_{\gamma}}{\sigma\hbar}}\rangle=
\expp^{-BT_{\text{loop}}/T_H} \; ,
\end{equation}
while the contribution from the $l$ orbit stretches in an $l$-encounter region is
\begin{equation} 
\langle\expp^{\frac{\ci xlQ^{enc}_{\gamma}}{\sigma\hbar}}\rangle=\expp^{-B l^2 t_{enc}/T_H} \; .
\end{equation}
This means that the average over the parametric velocities is now given by
\begin{equation} \label{qaver2}
\langle\expp^{\frac{\ci xQ_{\gamma}}{\sigma\hbar}}\rangle =
\expp^{-B (T - \sum_\alpha l_\alpha t_{\text{enc}}^\alpha) /T_H}
\expp^{-B \sum_\alpha l_\alpha^2 t_{\text{enc}}^\alpha /T_H} \; .
\end{equation}

The contribution of all orbit pairs with encounter regions described by $\v$ then
have the form
\begin{equation} \label{sumvp}
K_{\v}(\tau,x) = \frac{1}{T_H} \sum_{(\gamma,\gamma')}^{\text{fixed} \, \v} 
|A_\gamma|^2 e^{\ci \Delta S_\gamma/\hbar} 
\expp^{\frac{\ci x Q_\gamma}{\sigma \hbar}}
\delta(T - T_\gamma) = N(\v) \kappa \tau \int d^{L-V} s \; d^{L-V} u \;
\frac{z_T(\s,\u)}{L} \expp^{\ci \s \u / \hbar} \; ,
\end{equation}
where 
\begin{equation} 
\frac{z_{T}(s,u)}{L}=  \frac{w_{T}(s,u)}{L}
\langle\expp^{\frac{\ci xQ_{\gamma}}{\sigma\hbar}}\rangle 
=\frac{\expp^{-BT/T_H}T(T-\sum_{\alpha}l_{\alpha}t_{enc}^{\alpha})^{L-1}
\prod_{\alpha}\expp^{-Bl_\alpha(l_\alpha-1)t_{enc}^{\alpha}/T_H}}{L!\Omega^{L-V}
\prod_{\alpha}t_{enc}^{\alpha}} \; .
\end{equation}
Here $\alpha$ labels the $V$ different encounters,
each being a $l_{\alpha}$-encounter. Again because of (\ref{intenc}) the only
terms that contribute in the semiclassical limit are those where the encounter times
in the numerator and denominator cancel exactly. As a first step we can expand the
exponentials as a power series up to first order
\begin{equation} 
\frac{z_{T}(s,u)}{L}=\frac{\expp^{-B T/T_H}T(T-\sum_{\alpha}l_{\alpha}t_{enc}^{\alpha})^{L-1}
\prod_{\alpha}(1-l_\alpha(l_{\alpha}-1) B t_{enc}^{\alpha}/T_H)}{L!\Omega^{L-V}
\prod_{\alpha}t_{enc}^{\alpha}} \; .
\end{equation}
To obtain a product of the $V$ different encounter times in the numerator
we can take $r$ of them from the product over $\alpha$ and $V-r$ of them
from the bracket with the exponent $L-1$. The corresponding coefficient is
obtained by  combinatorial considerations. Then we sum over all values of $r$
from $0$ to $V$, and the result is
\begin{equation} 
\frac{z_{T}(s,u)}{L} \; \Longrightarrow \;
\frac{\expp^{-B T/T_H}T}{L!\Omega^{L-V}}
\sum_{r=0}^V  
\frac{T^{L-V+r-1} B^r (L-1)! (-1)^V \prod l^{v_l} }{T_H^r (L-V-1+r)! \; r!} 
\sum_{\substack{\alpha_1,...,\alpha_r \\ \text{distinct}}}
(l_{\alpha_1}-1) \times \ldots \times (l_{\alpha_r}-1) \; .
\end{equation}
We insert this into (\ref{sumvp}), evaluate the integral with
formula (\ref{intenc}), and obtain
\begin{equation} \label{kvtau}
K_{\v}(\tau,x) =
\frac{\kappa N(\v) \frac{\expp^{-B \tau}}{L}
\sum_{r=0}^V  
\tau^{L-V+r+1} B^r (-1)^V \prod l^{v_l} }{(L-V-1+r)! \; r!} 
\sum_{\substack{\alpha_1,...,\alpha_r \\ \text{distinct}}}
(l_{\alpha_1}-1) \times \ldots \times (l_{\alpha_r}-1) \; .
\end{equation}

The contribution of orbits for the different types of encounters can be calculated
with this formula, and they are shown for orbit pairs with $L-V \leq 4$ in
table~\ref{orbittablepara}. The vectors $\v$ are represented in the form
$(2)^{v_2}(3)^{v_3} \ldots$ and the horizontal lines separate vectors $\v$ with
different value of $L-V$. The numbers $N(\v)$ can be calculated by combinatorial
methods \cite{Mue05}.

\begin{table}[htb]
\centering
\begin{tabular}{|c|c|c|c|c|c|}
\hline
$\v$ & $L$ & $V$ & $K_{\v}(\tau,x)/(\kappa N(\v))$ & $N(\v)$, no TRS& $N(\v)$, TRS\\
\hline
$(2)^{1}$&2&1&$-\expp^{-B\tau}\left(\tau^{2}+B\tau^{3}\right)$&-&1\\
\hline
$(2)^{2}$&4&2&$\expp^{-B\tau}\left(\tau^{3}+B\tau^{4}
+\frac{B^{2}\tau^{5}}{6}\right)$&1&5\\
$(3)^1$&3&1&$-\expp^{-B\tau}\left(\tau^{3}+B\tau^{4}\right)$&1&4\\
\hline
$(2)^{3}$&6&3&$-\expp^{-B\tau}\left(\frac{2\tau^{4}}{3}+\frac{2B\tau^{5}}{3}+
\frac{B^{2}\tau^{6}}{6}+\frac{B^{3}\tau^{7}}{90}\right)$&-&41\\
$(2)^{1}(3)^1$&5&2&$\expp^{-B\tau}\left(\frac{3\tau^{4}}{5}+\frac{3B\tau^{5}}{5}+
\frac{B^{2}\tau^{6}}{10}\right)$&-&60\\
$(4)^{1}$&4&1&$-\expp^{-B\tau}\left(\frac{\tau^{4}}{2}
+\frac{B\tau^{5}}{2}\right)$&-&20\\
\hline
$(2)^{4}$&8&4&$\expp^{-B\tau}\left(\frac{\tau^{5}}{3}
+\frac{B\tau^{6}}{3}+\frac{B^{2}\tau^{7}}{10}
+\frac{B^{3}\tau^{8}}{90}+\frac{B^{4}\tau^{9}}{2520}\right)$&21&509\\
$(2)^{2}(3)^1$&7&3&$-\expp^{-B\tau}\left(\frac{2\tau^{5}}{7}+\frac{2B\tau^{6}}{7}
+\frac{B^{2}\tau^{7}}{14}+\frac{B^{3}\tau^{8}}{210}\right)$&49&1092\\
$(2)^{1}(4)^1$&6&2&$\expp^{-B\tau}\left(\frac{2\tau^{5}}{9}+\frac{2B\tau^{6}}{9}
+\frac{B^{2}\tau^{7}}{30}\right)$&24&504\\
$(3)^{2}$&6&2&$\expp^{-B\tau}\left(\frac{\tau^{5}}{4}+\frac{B\tau^{6}}{4}
+\frac{B^{2}\tau^{7}}{20}\right)$&12&228\\
$(5)^{1}$&5&1&$-\expp^{-B\tau}\left(\frac{\tau^{5}}{6}
+\frac{B\tau^{6}}{6}\right)$&8&148\\
\hline
\end{tabular}
\caption{Contribution of different types of orbit pairs to the parametric form factor}
\label{orbittablepara}
\end{table}

To find the total contribution to the form factor we now multiply the middle column 
that contains $K_{\v}/(\kappa N(\v))$ by $\kappa$ and $N(\v)$, add the diagonal approximation
and sum over different $\v$. If we do that for all orbits pairs with $L-V\leq8$
for the case without time reversal symmetry ($\kappa=1$), we obtain the expansion
for the form factor in $\tau$ up to 9th order
\begin{equation}
K(\tau)=\expp^{-B\tau}\left[\tau+\frac{B^{2}\tau^{5}}{6}+\frac{B^{4}\tau^{9}}{120}
+ \ldots \right] \; .
\end{equation}
This agrees with the first three terms of the expansion (\ref{kgue2}) in the section
on RMT.
It is noticable that when summing over terms with the same value of $L-V$, that all terms
cancel apart from the highest order term from orbit pairs with only 2-encounters.  In fact we
will show this using a recurrence relation in appendix A. This allows us to calculate
the expansion of the form factor to all orders in $\tau$ which is done in the next
section.

For systems with time reversal symmetry ($\kappa = 2$) we sum over all contributions
with $L-V\leq6$, and obtain the expansion of the parametric form factor in $\tau$ up
to 7th order
\begin{align}
K(\tau) & = \expp^{-B\tau} \left[ 2 \tau - 2 \tau^2 - (2B-2) \tau^3 +
\left(2 B - \frac{8}{3} \right) \tau^4 \right.
\notag \\ &
+ \left( \frac{5B^2}{3} - \frac{8B}{3} + 4 \right) \tau^5-
\left( \frac{5B^2}{3} - 4 B + \frac{32}{5} \right) \tau^6 
\notag \\ &
\left. - \left( \frac{41 B^3}{45} - \frac{11 B^2}{5} + \frac{32 B}{5}-
\frac{32}{3} \right) \tau^7 + \ldots \right] \; .
\end{align}
This agrees with the expansion of the RMT result (\ref{kgoe}).

\section{Systems without time reversal symmetry}

In this section we derive the full expansion of the parametric form factor
for small $\tau$ for the case of systems without time reversal symmetry.
For this purpose we rewrite the expansion $K(\tau,x) = \tau \expp^{-B \tau}
+ \sum_{n=2}^\infty K_{\v}(\tau,x)$
with $K_{\v}(\tau,x)$ given in equation (\ref{kvtau}) in the following form
\begin{equation} \label{formfa}
K(\tau,x) = \tau e^{-B \tau} + e^{-B \tau} \sum_{n=2}^\infty
\sum_{r=0}^{n-1} K_{n,r} \tau^{n+r} B^r \; .
\end{equation}
The coefficients $K_{n,r}$ have the form
\begin{equation} \label{knr}
K_{n,r} = \frac{1}{(n-2)!} S_n[f_r(\v)] \; ,
\end{equation}
where 
\begin{equation} \label{def_sn}
S_n[f_r(\v)] = \sum_{\v}^{L-V+1=n} f_r(\v) \tilde{N}(\v) \; ,
\end{equation}
and the functions $f_r(\v)$ are given by
\begin{equation} \label{def_fr}
f_r(\v) = \frac{(L-V-1)!}{(L-V-1+r)! \; r!} 
\sum_{\substack{\alpha_1,...,\alpha_r \\ \text{distinct}}}
(l_{\alpha_1}-1) \times \ldots \times (l_{\alpha_r}-1) \; .
\end{equation}
The first few functions are
\begin{align} \label{firstf}
f_0(\v) & = 1 \; , \notag \\
f_1(\v) & = \frac{\sum_\alpha (l_\alpha - 1)}{L-V}
= \frac{\sum_k v_k (k-1)}{L-V} = 1 \; , \notag \\
f_2(\v) & = \frac{\sum_{\alpha_1,\alpha_2} (l_{\alpha_1} - 1)
(l_{\alpha_2} - 1) - \sum_\alpha (l_\alpha - 1)^2}{2 (L-V) (L-V+1)}
= \frac{(L-V)^2 - \sum_k v_k (k-1)^2}{2 (L-V) (L-V+1)} \; .
\end{align}

We need to evaluate the quantities $S_n[f_r(\v)]$ for $r<n$.
In appendix \ref{recurrence} it is shown that
$S_n[f_r(\v)]=0$ for $r<n-1$. Hence the only non-vanishing
terms in the expansion (\ref{formfa}) are those with $r=n-1$.

\begin{equation}
K(\tau,x) = \tau e^{-B \tau} + e^{-B \tau} \sum_{n=2}^\infty
K_{n,n-1} \tau^{2n-1} B^{n-1} \; .
\end{equation}
Since $r$ satisfies $r \le V$ we have $V \ge n-1$. Together with
the condition $L-V=n-1$ we find that $2V \ge L$. This is only 
satisfied for orbit pairs with $V$ 2-encounters for which
$\v = (2)^V$ and $L=2V$. The contribution of these orbit pairs
to the form factors can be calculated explicitly. 
We obtain from equations (\ref{knr}), (\ref{def_sn}) and (\ref{def_fr})
with $r=V=n-1$, $L=2V$, and $l_{\alpha}=2$ for all $\alpha$,
\begin{equation}
K_{n,n-1} = \frac{\tilde{N}(\v)}{(n-2)!} \frac{(L-V-1)!}{(L-V-1+r)!}
= \frac{(-1)^{n-1} 2^{n-1}}{(2n-2)!} N(\v) \; .
\end{equation}
The number $N(\v)$ can be obtained from an explicit formula that
has been derived for systems without time reversal symmetry in
\cite{Mue03b}. In our notation it has the form
\begin{equation}
{\cal N}(\v) = \frac{1}{L+1} \sum_{\v' \leq \v}
\frac{(-1)^{L'-V'} \, L'! \, (L-L')!}{
\prod_{k \geq 2} k^{v_k} \, v_k'! \, (v_k-v_k')!} \; .
\end{equation}
The notation $\v' \leq \v$ means that the sum runs over all integer
vectors $\v'$ whose components satisfy $0 \leq v_k' \leq v_k$ for all $k$.
Furthermore, $L' = L(\v')$ and $V' = V(\v')$.
In the case of vectors $\v$ of the form $(2)^V$ the only
non-vanishing component of $\v$ is $v_2=V$ and the
sum runs over all vectors with component $v_2'=m$ where $m=0, \ldots, V$.
The result is
\begin{equation}
{\cal N}(\v) = \frac{1}{2V+1} \sum_{m=0}^V
\frac{(-1)^m \, (2m)! \, (2V-2m)!}{
2^V \, m! \, (V-m)!} = \frac{(2V)!}{2^V (V+1)!} \; \frac{(-1)^V+1}{2} \; ,
\end{equation}
where the last equality can be found in \cite{PBM88}.
The expression vanishes if $V$ is odd or equivalently if $n$ is even.  
With $V=n-1$, $n$ odd we obtain $K_{n,n-1} = 1/n!$ and complete expansion
of the form factor is ($n=2k+1$)
\begin{equation}
K(\tau,x) = \tau e^{-B \tau} + e^{-B \tau} \sum_{k=1}^\infty
\frac{\tau^{4k+1} B^{2k}}{(2k+1)!} = \frac{\sinh(B \tau^2)}{B \tau} \expp^{-B \tau}
\end{equation}
in agreement with the RMT result (\ref{kgue1}).

\section{Conclusions}

This work is a continuation of recent developments in semiclassical
periodic orbit expansions. These methods have been applied to 
several spectral and transport quantities in order to demonstrate
the universality of quantum fluctuation statistics of chaotic systems.
We extended these ideas to obtain a semiclassical
expansion of the parametric spectral form factor $K(\tau,x)$ for
small $\tau$ in agreement with RMT. For the GUE case we showed
agreement for all orders, while for the GOE case we showed agreement
for the first seven terms in the expansion. These terms can
actually be obtained more quickly by the semiclassical method
than by an expansion of the double integral for the GOE result.

The main input that is needed in addition to the
semiclassical calculation of the spectral form factor $K(\tau)$
is the distribution of the parametric velocities of long orbits.
This is the distribution of the derivatives of the actions of
periodic orbits with respect to the parameter, which is commonly
assumed to be Gaussian.
In addition, when averaging over orbit pairs whose correlations
are described by a particular {\em structure}, one has to take
into account correlations between different parts of the
{\it same} trajectory. These correlations are due to almost identical
orbit stretches within the encounter regions.

The limitations of the semiclassical calculation are similar
to that for the spectral form factor. One main open point
concerns the region $\tau>1$. In this regime the random matrix
expressions for $K(\tau,x)$ have a different functional form. 
So far it is not known how to extend the semiclassical approach
to this regime except for a small region near $\tau=1$ \cite{BK96,OLM98}.

\appendix

\section{Recurrence relations} \label{recurrence}

In this appendix we show that the quantities $S_n[f_r(\v)]$,
defined in (\ref{def_sn}) and (\ref{def_fr}), vanish for $r \leq n-1$.
The function $f_r(\v)$ are defined in terms of a restricted
sum in which all summation indices are distinct.
As a first step this sum is expressed by unrestricted sums.
How to do this by a combinatorial sieving is discussed, for
example, in section 4 of reference \cite{RS96}. 

We first introduce some notation.
A set partition $\F$ of the set of integers $\{1,2,\ldots\,r\}$ is a 
decomposition of this set into disjoint subsets
$[F_1,\ldots,F_\nu]$. Then $|F_1| + \ldots + |F_\nu| = r$
where $|F_i|$ is the number of elements in the set $F_i$.
Let us define a generalization of the Kronecker delta-function
\begin{equation}
\delta^{\F}_{\alpha_1,\ldots,\alpha_r} = \begin{cases} 1 &
\text{if $\alpha_i=\alpha_j$ for all $i$ and $j$ such that $i,j \in F_k$ for some $k$,} \\
0 & \text{otherwise.} \end{cases}
\end{equation}
Then 
\begin{equation}
\sum_{\substack{\alpha_1,...,\alpha_r \\ \text{distinct}}} [\ldots]
= \sum_{\F} \mu(\F) \sum_{\alpha_1,\ldots,\alpha_r} 
\delta^{\F}_{\alpha_1,\ldots,\alpha_r} [\ldots] \; ,
\end{equation}
where the first sum of the right-hand side runs over all
set partitions of the set of $r$ integers, and the
corresponding M\"obius function is given by
\begin{equation}
\mu(\F) = \prod_{i=1}^{\nu} (-1)^{|F_i|-1} (|F_i|-1)! \; .
\end{equation}
If we apply this to the functions $f_r(\v)$ we obtain
\begin{equation}
f_r(\v) =  \frac{(L-V-1)!}{(L-V-1+r)! \; r!} \sum_{\F} \mu(\F) g_{\F}(\v) \; ,
\end{equation}
where
\begin{equation}
g_{\F}(\v) = \left( \sum_{k} v_{k} (k - 1)^{|F_1|} \right) \times
\ldots \times \left( \sum_{k} v_{k} (k - 1)^{|F_\nu|} \right) \; .
\end{equation}
The first few functions $f_r(\v)$ were given in equation (\ref{firstf}).

The expansion of the form factor $K(\tau)$ was evaluated
in \cite{MHBHA04,MHBHA05} by using recurrence relations for the
number of structures ${\cal N}(\v)$ corresponding to a
vector $\v$. These recurrence relations were obtained by
relating orbits with $L$ loops to orbits with $L-1$
loops by considering all possible ways of removing a loop
(i.e. letting its size shrink to zero).

For systems without time reversal symmetry the relevant
recurrence relation is
\begin{equation} \label{mainrec}
v_2 \, \tilde{N}(\v) + \sum_{k \geq 2} v_{k+1}^{[k,2\rightarrow k+1]}
k \tilde{N}(\v^{[k,2\rightarrow k+1]}) = 0 \; .
\end{equation}
Here the vector $\v^{[k,2\rightarrow k+1]}$ is obtained from
the vector $\v$ by decreasing the components $v_k$ and $v_2$
by one and increasing the component $v_{k+1}$ by one. Hence
$L(\v^{[k,2\rightarrow k+1]}) = L(\v) - k - 2 + (k+1) =
L(\v) - 1$ and $V(\v^{[k,2\rightarrow k+1]}) = V(\v) - 1$.

In order to obtain the coefficient of the form factor expansion
one has to sum over the numbers ${\cal N}(\v)$ for all vectors
for which $L(\v)-V(\v)+1=n$. The recurrence relation may be
used for this purpose, because one can show that for each $k$
\begin{equation}
\sum_{\v}^{L-V+1=n}  v_{k+1}^{[k,2\rightarrow k+1]}
h(\v^{[k,2\rightarrow k+1]}) =
\sum_{\v'}^{L'-V'+1=n}  v_{k+1}' h(\v') \; ,
\end{equation}
where $h(\v)$ is some function of $\v$. One condition is that
$v_1=v_1^{[k,2\rightarrow k+1]}=0$, because the vectors describe
encounter regions which contain at least two orbit stretches.
Summing the recurrence relation (\ref{mainrec}) over $\v$ yields
\begin{equation} \label{sn1}
0 = S_n[v_2 +  \sum_{k \geq 2} v_{k+1} k] = S_n[L-V] = (n-1) S_n[1] \; .
\end{equation}
This shows, for example, that all off-diagonal terms of the
form factor $K(\tau,0)$ vanish \cite{MHBHA04,MHBHA05}.

We want to show in the following that $S_n[g_{\F}(\v)]=0$ if $r<n-1$.
We consider first the case when the partition consists of only one
subset $F_1$ with $|F_1|=r$. Then $g_{\F}(\v)=g_r(\v)$ where
\begin{equation} \label{defgr}
g_r(\v) = \sum_k v_k (k-1)^r \; .
\end{equation}
We show that $S_n[g_r(\v)]=0$ if $r<n-1$ by induction. The statement
is true for $r=0$, because $S_n[1]=0$ by equation (\ref{sn1}). Now we
fix a value of $r<n-1$ and assume that the statement is true for all
smaller values of $r$. From the definition (\ref{defgr}) follows that
\begin{equation} \label{rela1}
g_r(\v^{[k,2 \rightarrow k+1]}) = g_r(\v) - h_r(k) \; , \quad h_r(k) = (k-1)^r - k^r + 1 \; .
\end{equation}
Points that will be important in the following are that $h_r(1)=0$ and
that $h_r(k)$ is given by a finite power series in $k$ whose highest order
term is $-r k^{r-1}$. 

Multiplying equation (\ref{mainrec}) by $g_r(\v)$ and using
relation (\ref{rela1}) we obtain 
\begin{equation}
0 = v_2 \, g_r(\v) \, \tilde{N}(\v)
+ \sum_{k \geq 2} v'_{k+1} k g_r(\v') \tilde{N}(\v') 
+ \sum_{k \geq 2} v'_{k+1} k h_r(k)   \tilde{N}(\v') \; ,
\end{equation}
where $\v' = \v^{[k,2\rightarrow k+1]}$.
In the last sum we can start the sum at $k=1$, because $h_r(1)=0$,
and then change the summation index $k \rightarrow k-1$.
After summing over all vectors $\v$ we obtain
\begin{align}
0 & = S_n[v_2 g_r(\v) + \sum_{k \geq 2} v_{k+1} k g_r(\v)
+ \sum_{k \geq 2} v_k (k-1) h_r(k-1) 
\notag \\
 & = S_n[(L-V) g_r(\v) - \sum_{k \geq 2} v_k r (k-1)^r + \ldots ] \; .
\end{align}
In the second line we used that $v_2 + \sum_{k \ge 2}^\infty v_{k+1} k 
= \sum_{l \ge 2} v_l (k-1) = L-V$, and we wrote only the highest order
term of $h_r(k-1)$. The lower order terms, denoted by the dots,
involve powers $(k-1)^m$ with $m<r$ and can be neglected due to
our induction assumption. Hence we find that
\begin{equation}
(n-r-1) S_n[g_r(\v)] = 0 \; ,
\end{equation}
so that indeed $S_n[g_r(\v)]=0$ if $r<n-1$. 
The proof for general $g_{\F}(\v)$ is very similar. We consider the
general form
\begin{equation}
g_{\F}(\v) = \prod_{i=1}^\nu g_{|F_i|}(\v) \; ,
\end{equation}
and we use again induction to prove that $S_n[g_{\F}]=0$ if $r<n-1$. The statement is true
for $r=0$, and we fix a value of $r$ and assume that it is true for all smaller values of $r$.
In order to use the recurrence relation (\ref{mainrec}) we note that
\begin{equation} \label{rela2}
g_{\F}(\v) = \prod_{i=1}^\nu 
(g_{|F_i|}(\v^{[k,2 \rightarrow k+1]})+ h_{|F_i|}(k)) \; .
\end{equation}
We multiply equation (\ref{mainrec}) by $g_{\F}(\v)$ and use
relation (\ref{rela2}) to obtain 
\begin{align}
0 = & v_2 \, g_{\F}(\v) \, \tilde{N}(\v) + 
\sum_{k \geq 2} v'_{k+1} k g_{\F}(\v') \tilde{N}(\v') 
\notag \\ &
+ \sum_{k \geq 2} v'_{k+1}
k [\prod_{i=1}^\nu (g_{|F_i|}(\v') 
+ h_{|F_i|}(k)) - \prod_{i=1}^\nu g_{|F_i|}(\v')]
\tilde{N}(\v') \; ,
\end{align}
where we added an additional term and subtracted it again. 
As before $\v' = \v^{[k,2\rightarrow k+1]}$.
In the second sum we can start the sum at $k=1$, because $h_i(1)=0$ for all $i$,
and then change the summation index $k \rightarrow k-1$.
After summing over all vectors $\v$ we obtain
\begin{align}
0 & = S_n[v_2 g_{\F}(\v) + \sum_{k \geq 2} v_{k+1} k g_{\F}(\v)
+ \sum_{k \geq 2} v_k (k-1) [\prod_{i=1}^\nu (g_{|F_i|}(\v) + h_{|F_i|}(k-1)) 
- \prod_{i=1}^\nu g_{|F_i|}(\v)]
\notag \\
 & = S_n[(L-V) g_{\F}(\v) + \sum_{k \geq 2} v_k (k-1) \sum_{j=1}^\nu
(-|F_j| (k-1)^{|F_j|-1}) \prod_{i \neq j} g_{|F_i|}(\v) + \ldots ] \; .
\end{align}
In the step from the first to the second line we expanded the first product,
inserted the power series for the functions $h_{|F_i|}(k-1)$ and wrote 
only those terms that do not vanish due to the induction assumption. We obtain
further
\begin{align}
0 & =  S_n[(L-V) g_{\F}(\v) - \sum_{j=1}^\nu |F_j| \; g_{|F_i|}(\v)
\prod_{i \neq j} g_{|F_i|}(\v) + \ldots ]
\notag \\
 & = (n-1-r) S_n[g_{\F}(\v)] \; ,
\end{align}
which concludes the proof that $S_n[g_{\F}(\v)]=0$ for $r<n-1$.

\section*{Acknowledgements}

The authors would like to thank EPSRC for financial support.

\end{document}